\documentclass[onecolumn,showpacs]{revtex4}

\usepackage{graphicx}
\usepackage{dcolumn}
\usepackage{bm}

\input epsf

\begin{document}

\title{\Large A New Variable Modified Chaplygin Gas Model Interacting with Scalar Field}

\author{\bf Writambhara Chakraborty$^1$\footnote{writam1@yahoo.co.in}
and Ujjal Debnath$^2$\footnote{ujjaldebnath@yahoo.com} }

\affiliation{$^1$Department of Mathematics, New Alipore College,
New Alipore, Kolkata- 700 053, India.\\
$^2$Department of Mathematics, Bengal Engineering and Science
University, Shibpur, Howrah-711 103, India. }

\date{\today}

\begin{abstract}
In this letter we present a new form of the well known Chaplygin
gas model by introducing inhomogeneity in the EOS. This model
explains $\omega=-1$ crossing. Also we have given a graphical
representation of the model using $\{r,s\}$ parameters. We have
also considered an interaction of this model with the scalar field
by introducing a phenomenological coupling function and have shown
that the potential decays with time.
\end{abstract}


\maketitle

Recent observations reveals [1, 2] that the present Universe is
subjected to an accelerated expansion, which can be explained in
terms of some new type of matter which violates the strong energy
condition $\rho+3p<0$. This type of matter is known as dark energy
[3-6], which has the cosmological constant to be a strong
candidate. However many models have been proposed to play the role
of the dark energy, Quintessence [7] or the scalar field being one
of the most favoured model because of its decaying potential term
dominating the kinetic term so as to generate enough pressure to
drive acceleration. Also one can try Chaplygin gas model [8] with
equation of state (EOS), $ p=-B/\rho, $ as it generates negative
pressure, where $p$~ and $\rho$ are respectively the pressure and
energy density and $B$ is a positive constant. Subsequently this
fluid has been modified to $p=-B/\rho^{\alpha}~~ with~~ 0\le
\alpha \le 1.$ and
\begin{equation}
p=A\rho-\frac{B}{\rho^{\alpha}} ~~~~\text{with}~~~~ 0\le \alpha
\le 1,~A,~B~\text{are~positive~ constants}.
\end{equation}
as generalized Chaplygin gas [9, 10] and modified Chaplygin gas
[11, 12] respectively. Modified Chaplygin gas can explain the
evolution of the Universe from radiation era to $\Lambda$CDM
model. Later inhomogeneity has been introduced in the above EOS
(1) by considering $B$ to be a function of the scale factor $a(t)$
[13, 14]. This assumption is reasonable since $B(a)$ is related to
the scalar potential if we take the Chaplygin gas as a
Born-Infeld scalar field [15].\\

Interaction models where the dark energy weakly interacts with the
dark matter have also been studied to explain the evolution of the
Universe. This models describe an energy flow between the
components. To obtain a suitable evolution of the Universe the
decay rate should be proportional to the present value of the
Hubble parameter for good fit to the expansion history of the
Universe as determined by the Supernovae and CMB data. A variety
of interacting dark energy models have been proposed and studied
for this purpose [16-19]. \\

In this letter we study a new model by considering both $A$ and
$B$ in the EOS (1) to be a function of the scale factor $a(t)$ and
thus introducing inhomogeneity in the EOS (1). We solve the EOS to
get the energy density and show that the we can explain the
evolution of the Universe suitably by choosing different values of
the parameters. We then consider an interaction between the fluid
and the scalar field by introducing a phenomenological interaction
term which describes the energy flow between them, thus showing
the effect of interaction in the evolution of the Universe. This
kind of interaction term has been
studied in ref. [20].\\

The metric of a spatially flat homogeneous and isotropic universe
in FRW model is

\begin{equation}
ds^{2}=dt^{2}-a^{2}(t)\left[dr^{2}+r^{2}(d\theta^{2}+sin^{2}\theta
d\phi^{2})\right]
\end{equation}

where $a(t)$ is the scale factor.\\

The Einstein field equations are

\begin{equation}
\frac{\dot{a}^{2}}{a^{2}}=\frac{1}{3}\rho
\end{equation}
and
\begin{equation}
\frac{\ddot{a}}{a}=-\frac{1}{6}(\rho+3p)
\end{equation}

where $\rho$ and $p$ are energy density and isotropic pressure
respectively (choosing $8\pi G=c=1$).\\

The energy conservation equation is

\begin{equation}
\dot{\rho}+3\frac{\dot{a}}{a}(\rho+p)=0
\end{equation}

Now, we extend the modified Chaplygin gas with equation of state
(1) such that $A$ and  $B$ are positive function of the
cosmological scale factor `$a$' (i.e., $A=A(a), B=B(a)$). Then
equation (3) reduces to,
\begin{equation}
p=A(a)\rho-\frac{B(a)}{\rho^{\alpha}} ~~~~\text{with}~~~~ 0\le
\alpha \le 1
\end{equation}

As we can see this is an inhomogeneous EOS [21] where the pressure
is a function of the energy density $\rho$ and the scale factor
$a(t)$. Also if
$\rho=\left(\frac{B(a)}{A(a)}\right)^{\frac{1}{1+\alpha}}$,
this model reduces to dust model, pressure being zero.\\

Now, assume $A(a)$ and $B(a)$ to be of the form
\begin{equation}
A(a)=A_{0}a^{-n}
\end{equation}
and
\begin{equation}
B(a)=B_{0}a^{-m}
\end{equation}

where $A_{0}$, $B_{0}$, $n$ and $m$ are positive constants. If
$n=m=0$, we get back the modified Chaplygin gas [12] and if $n=0$,
we get back variable modified Chaplygin gas model. Using equations
(5), (6), (7) and (8), we get the solution of $\rho$
 as,

\begin{equation}
\rho=a^{-3} e^{\frac{3 A_{0}
a^{-n}}{n}}\left[C_{0}+\frac{B_{0}}{A_{0}} \left(\frac{3 A_{0}
(1+\alpha)}{n}\right)^{\frac{3(1+\alpha)+n-m}{n}}\Gamma(\frac{m-3(1+\alpha)}{n},\frac{3
A_{0}(1+\alpha)}{n}a^{-n})\right]^{\frac{1}{1+\alpha}}
\end{equation}

where $\Gamma(a,x)$ is the upper incomplete gamma function and $C_{0}$ is an integration constant .\\

Now, considering $$\omega_{eff}=\frac{p}{\rho}$$ for this fluid,
we have,$$ \omega_{eff}=A_{0} a^{-n}-B_{0} a^{-\zeta}
e^{-\frac{3A_{0}(1+\alpha) a^{-n}}{n}}\left[C_{0}+
\left(\frac{3A_{0}(1+\alpha)}{n}\right)^{\frac{n-\zeta}{n}}
\frac{B_{0}}{A_{0}}\Gamma(\frac{\zeta}{n},\frac{3A_{0}(1+\alpha)a^{-n}}{n})\right]^{-1}$$
where $ \zeta=m-3(1+\alpha)$.\\
For small values of the scale
factor $a(t)$, $\rho$ is very large and
$$p=A\rho-\frac{B}{\rho^{\alpha}}\rightarrow A\rho$$ where
$A=A_{0}a^{-n}$ is a function of $a$, so that for small scale
factor we have very large pressure and energy densities.
Therefore initially
$$\frac{p}{\rho}=\omega_{eff}=A^{*} a^{-n} \le 1$$ where
$A^{*}$ is a constant, $$A^{*}=A_{0}.$$ If
$a={A_{0}}^{\frac{1}{n}}$, the Universe starts from stiff perfect
fluid, and if $a={3 A_{0}}^{\frac{1}{n}}$, the Universe starts
from radiation era. \\

\begin{figure}
\includegraphics[height=2.2in]{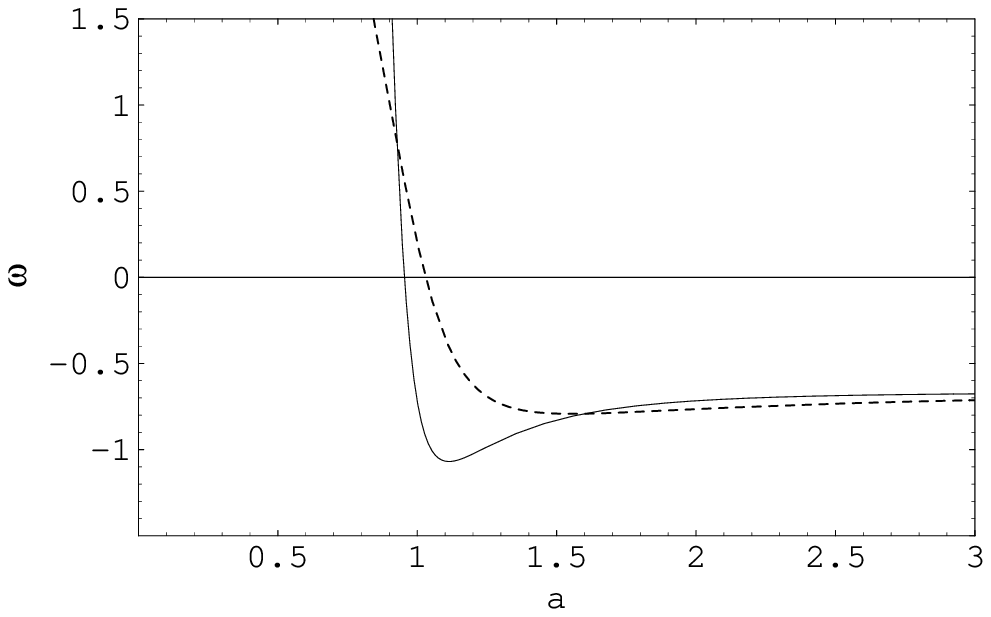}

Fig.1 \vspace{3mm}

\vspace{5mm} Fig. 1 shows the variation of $\omega_{eff}$ against
$a(t)$ for $ A_{0}=1, B_{0}=10, \alpha=1, m=2, C_{0}=1$ and
$n=3$(for dotted line), $n=10$(for the dark line). \hspace{14cm}
\vspace{4mm}

\end{figure}

Also for large values of the scale factor
$$p=A\rho-\frac{B}{\rho^{\alpha}}\rightarrow
-\frac{B}{\rho^{\alpha}}.$$ If $$\zeta=m-3(1+\alpha)<0$$ ( as we
know that upper incomplete Gamma function $\Gamma(a,x)$ exists for
$a<0$ ), the second term dominates and hence
$\omega_{_{eff}}\rightarrow -B^{*}a^{-\zeta}$, where
$$B^{*}=B_{0} {\lim_{a\rightarrow\infty}}{
e^{-\frac{3A_{0}(1+\alpha) a^{-n}}{n}}\left[C_{0}+
\left(\frac{3A_{0}(1+\alpha)}{n}\right)^{\frac{n-\zeta}{n}}
\frac{B_{0}}{A_{0}}\Gamma(\frac{\zeta}{n},\frac{3A_{0}(1+\alpha)a^{-n}}{n})\right]^{-1}}$$
( ${\lim_{a\rightarrow\infty}}{e^{-\frac{3A_{0}(1+\alpha)
a^{-n}}{n}}}\rightarrow 1$ and
${\lim_{a\rightarrow\infty}}{\Gamma(\frac{\zeta}{n},\frac{3A_{0}(1+\alpha)a^{-n}}{n})}\rightarrow$large
value, for $\zeta<0$ ). This will represent dark energy if
$a>\left(\frac{1}{3B^{*}}\right)^{\frac{1}{3(1+\alpha)-m}}$,
$\Lambda$CDM if
$a=\left(\frac{1}{B^{*}}\right)^{\frac{1}{3(1+\alpha)-m}}$ and
phantom dark energy if
$a>\left(\frac{1}{B^{*}}\right)^{\frac{1}{3(1+\alpha)-m}}$.
Therefore we can explain the evolution of the Universe till the
phantom era depending on the various values of the parameters. We
have shown a graphical representation of $\omega_{eff}$ in fig 1
for different values of the parameters. We can see from fig 1
that $\omega_{eff}$ starting from a large values decreases with
$a$ crosses $\omega=-1$ for some choices of the parameters.\\

\begin{figure}
\includegraphics[height=2.2in]{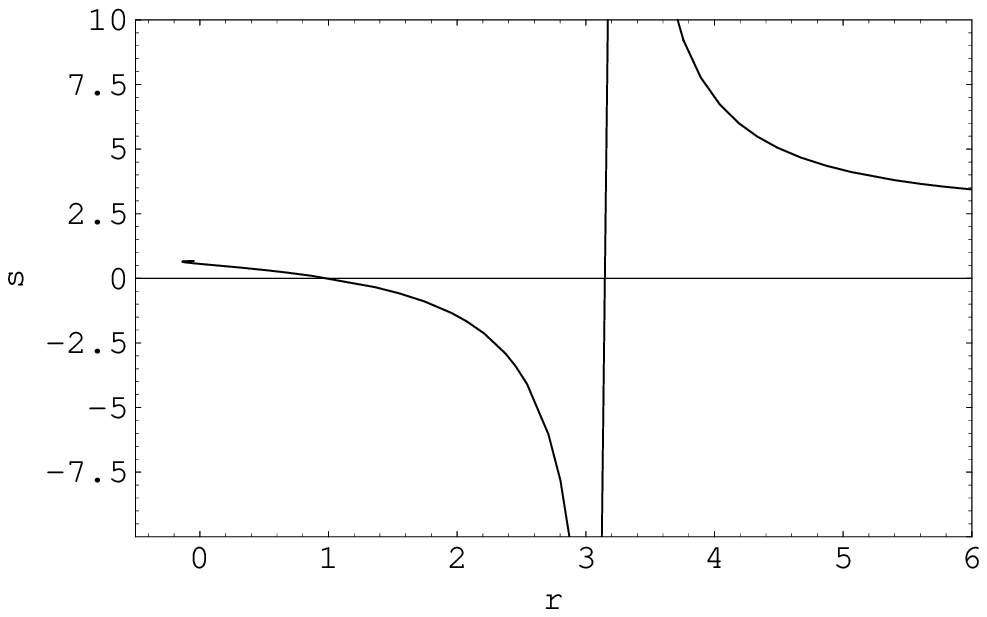}

Fig.2 \vspace{3mm}

\vspace{5mm} Fig. 2 shows the variation of $s$
 against $r$ for $ A_{0}=1, B_{0}=1, \alpha=\frac{1}{2}, m=3, n=2, C_{0}=1$. \hspace{14cm} \vspace{4mm}

\end{figure}

Since there are various candidates for the dark energy model, we
often face with the problem of discriminating between them, which
were solved by introducing statefinder parameters [22]. These
statefinder diagnostic pair i.e., $\{r,s\}$ parameters are of the
following form:
\begin{equation}
r=\frac{\dddot{a}}{aH^{3}}~~~~\text{and}~~~~s=\frac{r-1}{3\left(q-\frac{1}{2}\right)}
\end{equation}

where $H$ is the Hubble parameter and
$q~\left(=-\frac{a\ddot{a}}{\dot{a}^{2}}\right)$ is the
deceleration parameter. These parameters are dimensionless and
allow us to characterize the properties of dark energy in a model
independent manner. The statefinder is dimensionless and is
constructed from the scale factor of the Universe and its time
derivatives only. The parameter $r$ forms the next step in the
hierarchy of geometrical
cosmological parameters after $H$ and $q$.\\

Now, in our case,
\begin{equation}
H^{2}=\frac{\dot{a}^{2}}{a^{2}}=\frac{1}{3}\rho
\end{equation}
and
\begin{equation}
q=-\frac{\ddot{a}}{aH^{2}}=\frac{1}{2}+\frac{3}{2}\frac{p}{\rho}
\end{equation}

So from equation (10) we get

\begin{equation}
r=1+\frac{9}{2}\left(1+\frac{p}{\rho}\right)\frac{\partial
p}{\partial\rho}-\frac{3}{2}\frac{a}{\rho}\frac{\partial
p}{\partial a}
~,~~~~~s=\frac{2(r-1)}{9\left(\frac{p}{\rho}\right)}
\end{equation}

so that, solving we get,
\begin{equation}
r=1+\frac{9}{2}(1+y)(A_{0} a^{-n}+\alpha B_{0} a^{-m}
x)+\frac{3}{2}(n A_{0} a^{-n}-m B_{0}a^{-m}x)
~,~~~~~s=\frac{2(r-1)}{9y}
\end{equation}

where, $ y=\frac{p}{\rho}=A_{0}a^{-n}-B_{0}a^{-m} x$ and
$x=\rho^{-(1+\alpha)}$, $\rho$ is given by equation (9).\\

We have plotted the $ \{r, s\}$ parameters normalizing the
parameters and varying the scale factor $a(t)$. We can see that
the model starts from radiation era. Then we have a discontinuity
at the dust era (for radiation era: $s>0$ and $r>1$; dust era:
$r>1$ and $s\rightarrow \pm \infty$; $\Lambda$CDM: $r=1$, $s=0$;
phantom: $r<1$). The model reaches $\Lambda$CDM at $r=1,~s=0$ and
then crosses $\Lambda$CDM to represent phantom dark energy. This
model represents the phantom dark energy, whereas, Modified
Chaplygin Gas can explain the evolution of the Universe from
radiation to $\Lambda$CDM and Variable Modified Chaplygin gas
describes the evolution of the Universe from radiation to quiessence model.\\

Now we consider model of interaction between scalar field and the
new variable modified Chaplygin Gas model, through a
phenomenological interaction term. Keeping into consideration the
fact that the Supernovae and CMB data determines that decay rate
should be proportional to the present value of the Hubble
parameter. This interaction term describes the energy flow between
the two fluids. We have considered a scalar field to couple with
the New variable modified Chaplygin gas
given by EOS (6), (7) and (8). \\

Therefore now the conservation equation becomes

\begin{equation}
\dot{\rho}_{tot}+3\frac{\dot{a}}{a}(\rho_{tot}+p_{tot})=0
\end{equation}
so that the equations of motion of the the new fluid and scalar
field read,
\begin{equation}
\dot{\rho}+3H(\rho+p)=-3H\rho\delta
\end{equation}
and
\begin{equation}
\dot{{\rho}}_{\phi}+3H({\rho}_{\phi}+p_{\phi})=3H\rho\delta
\end{equation}
( $\delta$ is a constant ).\\

Where the total energy density and pressure of the universe are
given by,
\begin{equation}
\rho_{tot}=\rho+\rho_{\phi}
\end{equation}
and
\begin{equation}
p_{tot}=p+p_{\phi}
\end{equation}

where $\rho$ and $p$ are the energy density and pressure of the
extended modified Chaplygin gas model given by equations (6), (7),
(8), (9) and $\rho_{\phi}$ and $p_{\phi}$ are the energy density
and pressure due to the scalar field given by,
\begin{equation}
\rho_{\phi}=\frac{{\dot{\phi}}^{2}}{2}+V(\phi)
\end{equation}
and
\begin{equation}
p_{\phi}=\frac{{\dot{\phi}}^{2}}{2}-V(\phi)
\end{equation}
where, $V(\phi)$ is the relevant potential for the scalar field
$\phi$.\\

Thus the field equations become
\begin{equation}
\frac{\dot{a}^{2}}{a^{2}}=\frac{1}{3}\rho_{tot}
\end{equation}
and
\begin{equation}
\frac{\ddot{a}}{a}=-\frac{1}{6}(\rho_{tot}+3p_{tot})
\end{equation}

Solving the equations we get the solution for $\rho$ as
\begin{equation}
\rho=a^{-3(1+\delta)} e^{\frac{3 A_{0}
a^{-n}}{n}}\left[C_{0}+\frac{B_{0}}{A_{0}} \left(\frac{3 A_{0}
(1+\alpha)}{n}\right)^{\frac{3(1+\alpha)(1+\delta)+n-m}{n}}\Gamma(\frac{m-3(1+\alpha)(1+\delta)}{n},\frac{3
A_{0}(1+\alpha)}{n}a^{-n})\right]^{\frac{1}{1+\alpha}}
\end{equation}
where $C_{0}$ is an integration constant.\\

Further substitution in the above equations give,

\begin{equation}
V(\phi)=3H^{2}+\dot{H}+\frac{p-\rho}{2}
\end{equation}
To get an explicit form of the energy density and the potential
corresponding to the scalar field we consider a power law
expansion of the scale factor $a(t)$ as,
\begin{equation}
a=t^{\beta}
\end{equation}
so that, for $\beta>1$ we get accelerated expansion of the
Universe thus satisfying the observational constrains. If
$\beta=1$ or $\beta<1$ we get constant and decelerated expansion
respectively.\\

Using equations (18), (22) and (26), we get,
\begin{equation}
\rho_{\phi}=\frac{3 \beta^{2}}{t^{2}}-\rho
\end{equation}
where $\rho$ is given by equation (25) along with (27). Since
$\rho_{\phi}$ is always positive, so we may have at least for some
range of the values of the free parameters.\\

Also the potential takes the form,
\begin{equation}
V=\frac{3\beta^{2}-\beta}{t^{2}}+\frac{p-\rho}{2}
\end{equation}
The graphical representation of $V$ against time is shown in
figure 3 normalizing the parameters. We see that the potential
decays with time.\\

\begin{figure}
\includegraphics[height=1.7in]{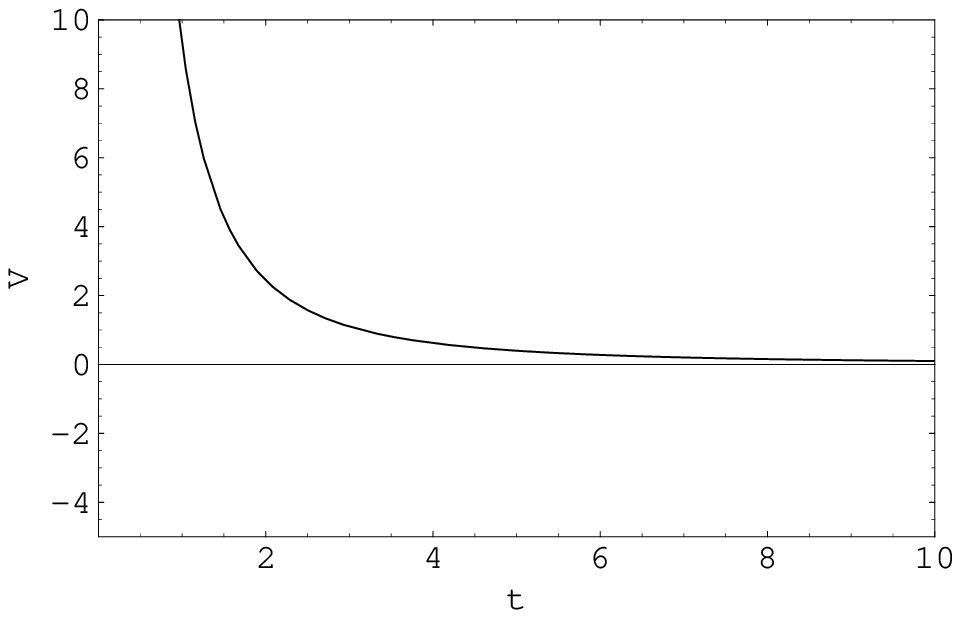}

Fig.3 \vspace{5mm}

\vspace{5mm} Fig. 3 shows the variation of $V$
 against $t$ for $ A_{0}=1, B_{0}=1, \alpha=\frac{1}{2}, m=3, n=2, C_{0}=0,~\delta=0.01$. \hspace{14cm} \vspace{4mm}

\end{figure}

Here we present a new variable modified Chaplygin gas model which
is an unified version of the dark matter and the dark energy of
the Universe. It behaves like dark matter at the initial stage and
later it explains the dark energy of the Universe. Unlike the
Generalized or Modified Chaplygin gas model, it can explain the
evolution of the Universe at phantom era depending on the
parameters. Also we have calculated the $\{r,s\}$ parameters
corresponding to this model. Normalizing the parameters such that
$m-3(1+\alpha)<0$, show the diagrammatical representation of
$\{r,s\}$ for our model (in Fig.2), varying the scale factor. We
see that starting from the radiation era it crosses $\omega=-1$
and extends till phantom era. Also we can see that the
deceleration parameter starting from a positive point becomes
negative, indicating deceleration initially and acceleration at
later times. Again we have considered an interaction of this fluid
with that of scalar field by introducing a phenomenological
coupling term, so that there is a flow of energy between the field
and the fluid which decays with time, as in the initial stage the
fluid behaves more like dark matter and the field that of dark
energy, whereas in the later stage both explain the dark energy
present in the Universe. In Fig.3, we have shown the nature of the
potential by considering a power law expansion of the Universe to
keep the recent observational support of cosmic acceleration, and
we see that the potential
decays with time.\\\\

{\bf Acknowledgement:}\\

The authors are thankful to IUCAA, India for warm hospitality
where part of the work was carried out. Also UD is thankful to
UGC, Govt. of India for providing research project grant (No. 32-157/2006(SR)).\\

{\bf References:}\\
\\
$[1]$  N. A. Bachall, J. P. Ostriker, S. Perlmutter and P. J.
Steinhardt, {\it Science} {\bf 284} 1481 (1999).\\
$[2]$ S. J. Perlmutter et al, {\it Astrophys. J.} {\bf 517} 565
(1999).\\
$[3]$ V. Sahni and A. A. Starobinsky, {\it Int. J. Mod. Phys. A}
{\bf 9} 373 (2000).\\
$[4]$ P. J. E. Peebles and B. Ratra, {\it Rev. Mod. Phys.} {\bf
75} 559 (2003).\\
$[5]$ T. Padmanabhan, {\it Phys. Rept.} {\bf 380} 235 (2003).\\
$[6]$ E. J. Copeland, M. Sami, S. Tsujikawa, {\it Int. J. Mod.
Phys. D} {\bf  15} 1753 (2006).\\
$[7]$ S. Das, N. Banerjee, {\it Gen. Rel. Grav.} {\bf 38} 785
(2006).\\
$[8]$ A. Kamenshchik, U. Moschella and V. Pasquier, {\it Phys.
Lett. B} {\bf 511} 265 (2001); V. Gorini, A. Kamenshchik, U.
Moschella and V. Pasquier, {\it gr-qc}/0403062.\\
$[9]$ V. Gorini, A. Kamenshchik and U. Moschella, {\it Phys. Rev.
D} {\bf 67} 063509 (2003); U. Alam, V. Sahni , T. D. Saini and
A.A. Starobinsky, {\it Mon. Not. Roy. Astron. Soc.} {\bf 344}, 1057 (2003).\\
$[10]$ M. C. Bento, O. Bertolami and A. A. Sen, {\it Phys. Rev. D}
{66} 043507 (2002).\\
$[11]$ H. B. Benaoum, {\it hep-th}/0205140.\\
$[12]$ U. Debnath, A. Banerjee and S. Chakraborty, {\it Class.
Quantum Grav.} {\bf 21} 5609 (2004).\\
$[13]$ Z. K. Guo and Y. Z. Zhang, {\it Phys. Lett. B} {\bf 645} 326 (2007), {\it astro-ph}/0506091.\\
$[14]$ G. Sethi, S. K. Singh, P. Kumar, D. Jain and A. Dev, {\it
Int. J. Mod. Phys. D} {\bf 15} 1089 (2006); Z. K. Guo and Y. Z.
Zhang, {\it astro-ph}/0509790.\\
$[15]$ M.C. Bento, O. Bertolami and A.A. Sen, {\it Phys. Lett. B}
{\bf 575} 172 (2003).\\
$[16]$ M. S. Berger, H. Shojaei, {\it Phys. Rev. D} {\bf 74}
043530 (2006).\\
$[17]$ R.-G. Cai, A. Wang, {\it JCAP}
{\bf 03} 002 (2005).\\
$[18]$ W. Zimdahl, {\it Int. J. Mod. Phys. D} {\bf 14}2319
(2005)\\
$[19]$ B. Hu, Y. Ling, {\it Phys. Rev. D} {\bf 73} 123510
(2006).\\
$[20]$ R.-G. Cai, A. Wang, {\it JCAP} {\bf 03} 002 (2005).\\
$[21]$ I. Brevik, O. G. Gorbunova, A. V. Timoshkin, {\it Eur.
Phys. J. C} {\bf 51} 179 (2007).\\
$[22]$ V. Sahni, T. D. Saini, A. A. Starobinsky and U. Alam, {\it
JETP Lett.} {\bf 77} 201 (2003).\\

\end{document}